\documentclass{mem}
\usepackage{natbib}\usepackage{txfonts}\usepackage{balance}
\usepackage{graphicx}
\usepackage[a4paper,breaklinks,dvipdfm]{hyperref}
\usepackage{color}
\usepackage{booktabs}  
\usepackage{tabularx} 

\begin{document}

\newcommand{\todo}[1]{\textbf{\color{red}TODO: #1}}

\newcommand{\gcc}{\ensuremath{\mathrm{g} \, \mathrm{cm}^{-3}}}
\newcommand{\cms}{\ensuremath{\mathrm{cm\, s^{-1}}}}
\newcommand{\kms}{\ensuremath{\mathrm{km\, s^{-1}}}}
\newcommand{\cm}{\ \mathrm{cm}}
\newcommand{\mm}{\ensuremath{\mathrm{mm}}}
\newcommand{\foe}{\ensuremath{\mathrm{10^{51}\ erg\ }}}
\newcommand{\nuc}[2]{\ensuremath{\mathrm{^{#1}#2}}}
\newcommand{\ions}[2]{#1\,{\sc #2}}
\newcommand{\ye}{\ensuremath{Y_\mathrm{e}}}
\newcommand{\msun}{\ensuremath{\mathrm{M}_\odot}}
\newcommand{\gccm}{\ensuremath{\mathrm{g} \, \mathrm{cm}^{-3}}}
\newcommand{\mch}{\ensuremath{M_\mathrm{Ch}}}
\def\lesssim{\mathrel{\hbox{\rlap{\hbox{\lower4pt\hbox{$\sim$}}}\hbox{$<$}}}}
\let\la=\lesssim
\def\gtrsim{\mathrel{\hbox{\rlap{\hbox{\lower4pt\hbox{$\sim$}}}\hbox{$>$}}}}
\let\ga=\gtrsim

\title{
  Simulating the observed diversity of Type Ia supernovae
}

\subtitle{Introducing a model data base}

\author{
  M.\,Kromer\inst{1,2},
  S.T.\,Ohlmann\inst{1},  
  \and F.K.\,R\"opke\inst{1,2}
}

\institute{
 $^{1}$Heidelberger Institut f\"{u}r Theoretische Studien, 
        Schloss-Wolfsbrunnenweg 35, D-69118 Heidelberg, Germany\\
  $^{2}$Institut f{\"u}r Theoretische Astrophysik am Zentrum f{\"u}r Astronomie der Universit{\"a}t Heidelberg, 
       Philosophenweg 12, 
       D-69120 Heidelberg, Germany\\
\email{markus.kromer@h-its.org}
}

\authorrunning{Kromer}

\titlerunning{Type Ia supernova models}

\abstract{Despite the importance of Type Ia supernovae (SNe~Ia) for
  modern astrophysics, their detailed mechanism is still not fully
  understood. In this contribution, we present recent findings from
  numerical explosion models in the context of the observed diversity
  of SNe~Ia and we discuss how these models can help to shed light on
  the explosion mechanism and the progenitor stars of SNe~Ia. In
  addition, we introduce the Heidelberg Supernova Model Archive
  (HESMA), a new online data base where we provide integrated isotopic
  abundances and radially averaged ejecta profiles and synthetic
  observables for a wide range of state-of-the-art explosion models.

}

\maketitle{}

\section{Introduction}

Type Ia supernovae (SNe~Ia) are characterised by prominent absorption
features of \ions{Si}{ii} and the absence of H features in their
maximum-light spectra \citep[e.g.,][]{filippenko1997a}. An empirical
relation between their peak brightness and light-curve width
\citep{phillips1993a} makes them standardisable candles. Combined with
their large intrinsic luminosity this makes them important tools for
cosmological distance measurements \citep{riess1998a,
  perlmutter1999a}.

There is a wide consensus that SNe~Ia originate from thermonuclear
explosions of carbon--oxygen (CO) white dwarfs (WD) in interacting
binary systems \citep{hoyle1960a} and that their light curves are
powered by the radioactive decay of \nuc{56}{Ni} and its daughter
nucleus \nuc{56}{Co} \citep{colgate1969a}. However, the actual
progenitor stars are still elusive \citep[e.g.,][for a
review]{maoz2014a} and details of the explosion mechanism not fully
understood \citep[e.g.,][]{hillebrandt2000a}.

In the absence of direct progenitor detections, theoretical models are
one promising way to further our understanding of SNe~Ia. To this end,
we have developed a suite of numerical codes for a detailed simulation
of the thermonuclear burning, nucleosynthesis processes and the
propagation of radiation through the resulting explosion ejecta in
SNe~Ia. Employing this simulation pipeline to explore the predictions
of various progenitors scenarios and explosion mechanisms in the
context of observed SN~Ia properties allows us to derive constraints
on certain scenarios \citep[see e.g.,][]{roepke2012a}.

In this contribution, we briefly present simulations for various
progenitor models and discuss their properties in the context of the
observed diversity of SNe~Ia
(Sections~\ref{sec:MCh_models}-\ref{sec:mergers}). In
Section~\ref{sec:hesma}, we introduce the Heidelberg Supernova Model
Archive, which provides isotopic abundances and synthetic observables
for a large variety of explosion models to the community.

\section{Chandrasekhar-mass models}
\label{sec:MCh_models}

In the canonical model of SNe~Ia, a CO WD accretes H-rich material
from a companion star \citep[an evolved main-sequence star or a red
giant,][]{whelan1973a}. When it nears the Chandrasekhar stability
limit (\mch) pycnonuclear reactions ignite in the WD core. After a
simmering phase that lasts for several hundred years a thermonuclear
runaway occurs in the electron degenerate WD matter \citep[see
e.g.,][for a detailed description]{hillebrandt2000a}. The
thermonuclear burning can either proceed as a subsonic deflagration
(mediated by thermal conduction) or as a shock-driven, supersonic
detonation.

For a \mch\ WD the initial density is so high (a few times
$10^9\,\gcc$) that a detonation yields only iron-group elements
(IGEs), excluding this channel for SNe~Ia since their spectra indicate
the presence of significant amounts of intermediate-mass elements
(IMEs, e.g., Si, S, Ca) in the ejecta \citep{arnett1969a}. In
contrast, in a deflagration the WD can react to the energy release
from the burning and expands. This, in turn, decreases the burning
efficiency. Recent 3D models indicate that deflagrations yield at most
$\sim0.4\,\msun$ of \nuc{56}{Ni} \citep{fink2014a}, which falls short
to explain the peak brightness of the bulk of normal SNe~Ia
(observations indicate the need of $\sim0.5-0.6\,\msun$ of
\nuc{56}{Ni} to power their peak brightness \citealt{scalzo2014a}).
However, it has been shown that weak deflagrations that fail to
completely unbind the progenitor WD explain the observed properties of
a peculiar sub-class of faint SNe~Ia, the so called Type Iax
supernovae \citep[e.g.,][]{jordan2012b, kromer2013a, magee2016a}.

One way to produce normal SNe~Ia from \mch\ WDs is the delayed
detonation scenario. Here, the burning starts as a deflagration, which
allows the WD to expand before the burning mode switches to a
detonation. This may happen either by a spontaneous
deflagration-to-detonation transition \citep[DDT,
e.g.,][]{khokhlov1991a, seitenzahl2013a}, or the convergence of
macroscopic fluid flows like in the gravitationally-confined
detonation model \citep[GCD, e.g.,][]{plewa2004a, seitenzahl2016a}. In
particular 1D DDT models have been very successful in explaining the
observed properties of normal SNe~Ia in the past
\citep[e.g.,][]{hoeflich1996a, blondin2013a}. However, recent 3D DDT
models have problems to reproduce the observed colours and the
light-curve width-luminosity relation of SNe~Ia \citep{sim2013a}. The
reason for this seems to be turbulent mixing during the initial
deflagration phase that produces ejecta with significant amounts of
stable IGEs outside the \nuc{56}{Ni} region. In contrast, in 1D models
turbulence is suppressed, leading to a concentration of stable IGEs in
the centre of the ejecta.

\section{Sub-Chandrasekhar-mass models}
\label{sec:subMCh_models}

If a sub-Chandrasekhar-mass CO WD accretes He from a companion star
(either a He-rich donor or a He WD), a detonation may be triggered in
the accumulated He layer when the right conditions are met
\citep{taam1980a,shen2010a}.  In this case, an ensuing detonation of
the underlying sub-\mch\ CO core seems to be almost unavoidable
\citep{fink2007a, moll2013a}.  This so called double detonation
mechanism \citep{iben1987a} has the potential to produce a wide range
of \nuc{56}{Ni} masses \citep[e.g.,][]{fink2010a, woosley2011b} that
covers faint and normally bright SNe~Ia. The amount of \nuc{56}{Ni}
produced is thereby directly proportional to the initial mass of the
WD.  However, IGE-rich burning ashes of the He shell detonation leave
characteristic imprints on synthetic spectra that are not in agreement
with the observed spectra of SNe~Ia \citep[e.g.,][]{hoeflich1996a,
  kromer2010a}.

The final burning products of the He shell detonation depend very
sensitively on the initial conditions in that shell. For example, an
admixture of C nuclei to the He shell can suppress the production of
IGEs leading to much better agreement with the observed properties of
SNe~Ia \citep{kromer2010a}. In the limit of bare sub-\mch\ CO WDs,
studied by \citet{shigeyama1992a} and \citet{sim2010a}, centrally
ignited detonations qualitatively reproduce observed trends of normal
SNe~Ia like the light curve width--luminosity relation and the
evolution of the Si line ratio from faint to bright SNe~Ia. Lower mass
sub-\mch\ WDs that produce only $\sim0.1\,\msun$ of \nuc{56}{Ni} may
be good candidates to explain the sub-luminous 1991bg-like SNe~Ia
\citep{hachinger2009a}.

\section{Violent mergers}
\label{sec:mergers}

Another progenitor channel for SNe~Ia presents the merger of
two CO WDs that together exceed the Chandrasekhar mass limit
\citep{webbink1984a}. From a theoretical point of view, however, this
channel was heavily debated for a long time, since many simulations had
shown that WD mergers undergo an accretion induced collapse rather
than a thermonuclear explosion \citep[e.g.,][]{nomoto1985a, saio2004a,
  shen2012a}. All these simulations assumed that the lower mass WD is
disrupted and then slowly accreted onto the primary (i.e., more
massive) WD.

More recent simulations \citep[e.g.,][]{pakmor2011b, pakmor2013a,
  tanikawa2015a} indicate that a detonation may form when the two WDs
come in contact and the secondary WD is still intact. In this violent
merger scenario the system is disrupted by a thermonuclear
explosion. Since the mass of the primary WD is below \mch\ at the time
of explosion, the violent merger scenario is basically a sub-\mch\
explosion scenario (even if the total ejecta mass can be larger than
\mch, if the secondary WD is disrupted by the explosion). The violent
merger scenario inherits all the positive properties of detonations of
bare sub-\mch\ CO WDs. In particular, it can cover a wide range of
peak brightness and the flux spectra qualitatively match those of
observed SNe~Ia \citep[][]{pakmor2012a, moll2014a}. However, the
interaction of the ejecta with the secondary WD introduce strong
asymmetries leading to significant continuum polarisation in the
spectra, which is not observed in normal SNe~Ia \citep{bulla2016a}.

Nonetheless, violent mergers with lower mass primaries ($\sim$$0.9\,\msun$)
show remarkably similar observables to a group of slowly-evolving
faint SNe~Ia similar to SN~2002es \citep{kromer2013b, kromer2016a}. It
will be interesting to obtain spectropolarimetry for a future event of
this class to see whether a violent merger model is compatible.

\section{Heidelberg Supernova Model Archive (HESMA)}
\label{sec:hesma}

Following repeated requests by the community for nucleosynthetic
yields, ejecta profiles and synthetic observables of our explosion
models, we have decided to set up a public model archive, the
Heidelberg Supernova Model Archive
(HESMA)\footnote{\url{https://hesma.h-its.org}}. We hope this
facilitates further work, e.g., in galactic chemical evolution
and the interpretation of observed SN spectra.

\begin{figure*}[t!]
  \resizebox{\hsize}{!}{\includegraphics{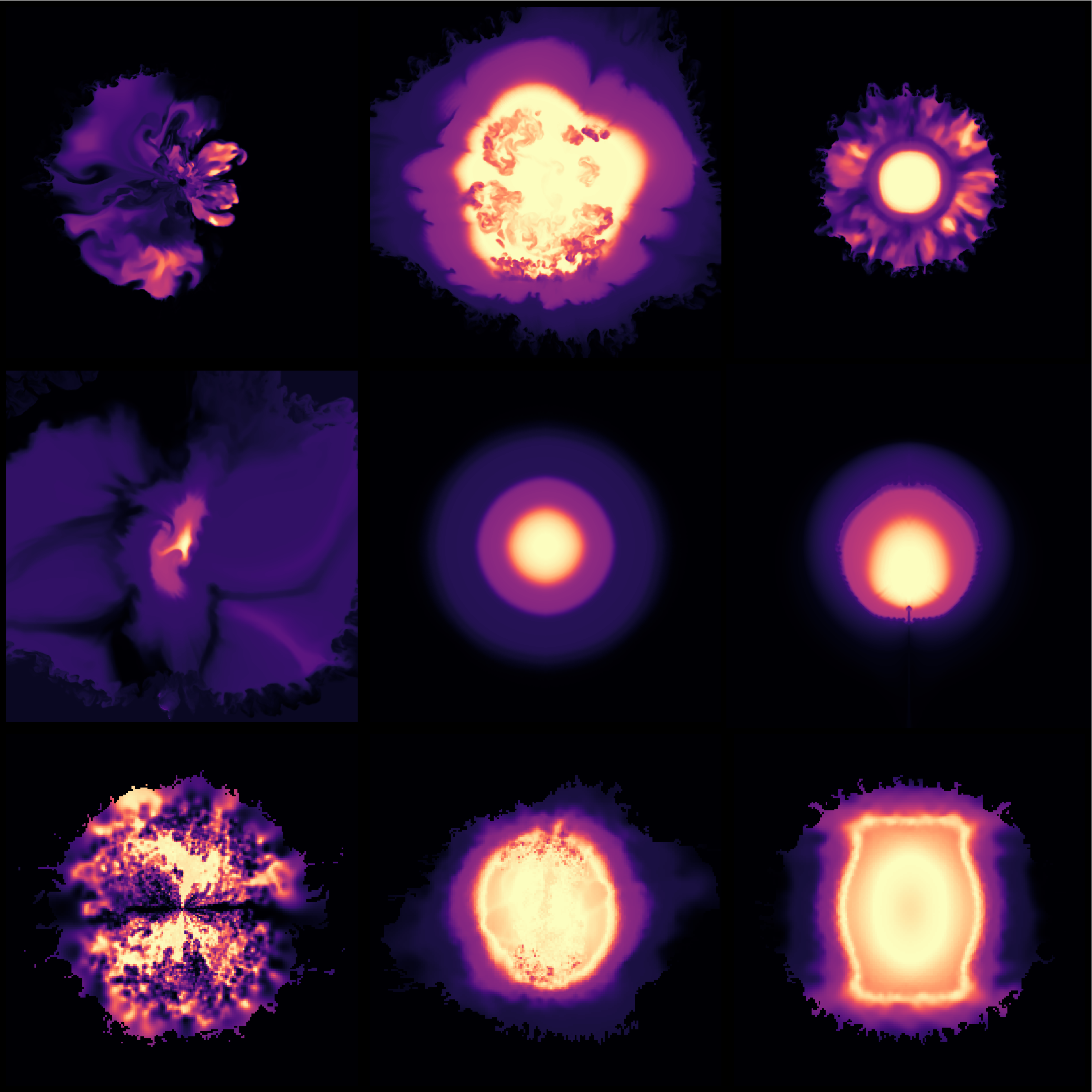}}
  \caption{\footnotesize Displayed are slices of the mean atomic number at the
    end of the hydrodynamics simulation for a variety of models from HESMA\@. The
    top row shows Chandrasekhar-mass explosions (from left to right:
    deflagration, delayed detonation and gravitationally confined detonation),
    the middle row shows explosions of a sub-Chandrasekhar-mass primary WD
    (from left to right: violent merger, pure detonation, double detonation),
    and the bottom row shows super-Chandrasekhar-mass explosions (from left to
    right: deflagration, delayed detonation, detonation).}
  \label{fig:hesma_models}
\end{figure*}

Currently, the archive comprises about 70 (multi-)dimensional
explosion models from the SN~Ia group formerly based at the Max Planck
Institute for Astrophysics in Garching (MPA) and now spread to
Heidelberg, Belfast and Canberra.  The models cover a wide range of
progenitor scenarios and explosion mechanisms.  For an overview of the
included models see Table~\ref{tab:hesma}.  A graphical representation
of a subset of the models is shown in Figure~\ref{fig:hesma_models}.

All the explosion simulations were performed with the hydrodynamics
code LEAFS \citep{reinecke2002b}.  Simulations of the merging process
in the violent merger scenario were conducted using GADGET
\citep{pakmor2012b}.  Detailed isotopic abundances were obtained from
nucleosynthesis calculations employing a large 384-isotope network on
thermodynamic trajectories of tracer particles \citep{travaglio2004a}.
Finally, synthetic observables were derived with the Monte Carlo
radiative transfer code ARTIS \citep{kromer2009a}.

\begin{table*}
  \caption{Overview of simulations presently available in HESMA.}
  \label{tab:hesma}
    \begin{tabularx}{\textwidth}{lllX}
      \toprule
      Progenitor & Explosion mechanism & Number & Source \\
      \midrule
      sub-\mch\ CO  & double detonation  & 12 & \citet{fink2010a,kromer2010a,sim2012a}\\
      sub-\mch\ CO  & core detonation    &  7 & \citet{sim2010a,marquardt2015a}\\
      sub-\mch\ CO  & shell detonation   &  2 & \citet{sim2012a}\\
      sub-\mch\ ONe & core detonation    &  5 & \citet{marquardt2015a}\\
      \mch\ CO      & deflagration       & 14 & \citet{kromer2013a,fink2014a}\\
      \mch\ CO      & spontaneous DDT    & 20 & \citet{seitenzahl2013a,sim2013a,ohlmann2014a}\\
      \mch\ CO      & GCD                &  1 & \citet{seitenzahl2016a}\\
      \mch\ hybrid  & deflagration       &  1 & \citet{kromer2015a}\\
      CO+CO merger  & detonation         &  4 &        
          \citet{pakmor2010a,pakmor2012a,kromer2013b,kromer2016a}\\
      super-\mch\ CO& various            &  5 & \citet{fink2010b,hillebrandt2013a}\\
      \bottomrule
    \end{tabularx}
\end{table*}

In HESMA, we provide integrated isotopic abundances as well as
radially averaged density and isotopic abundance profiles of the
explosion ejecta after they achieved homologous expansion. We also
offer angle-averaged UVOIR bolometric light curves and gamma-ray and
optical spetral time series. Bolometric light curves are tabulated
from $\sim 6$ to 80\,d past explosion. Gamma-ray spectra for selected
models are available from $\sim 6$ to 100\,d past explosion. Optical
($\lambda\in[3500,9500]\,\AA$) spectral time series are available from
$\sim 6$ to 40\,d past explosion and will also be ingested to the
Weizmann Interactive Supernova data Repository (WISeREP,
\citealt{yaron2012a}).

\section{Conclusions}

We presented a brief overview of a variety of state-of-the-art
explosion models of CO WDs and discussed their properties in the
context of the observed diversity of SNe~Ia.

We also introduced HESMA, the Heidelberg Supernova Model Archive that
comprises data for about 70 models on the hydrodynamical structure of
the ejecta, on nucleosynthetic abundances as well as on synthetic
observables: bolometric light curves and spectral time series in the
optical and gamma rays.

\begin{acknowledgements}
  We thank all our colleagues from the former MPA and W\"urzburg SN~Ia
  groups for the effort they put into the creation of the HESMA
  models. Specifically, we want to mention Michael Fink, R\"udiger
  Pakmor, Ivo Seitenzahl, Stuart Sim, and, in particular, Wolfgang
  Hillebrandt who brought the group to being in the first place.  The
  authors gratefully acknowledge the Gauss Centre for Supercomputing
  (GCS) for providing computing time through the John von Neumann
  Institute for Computing (NIC) on the GCS share of the supercomputer
  JUQUEEN \citep{stephan2015a} at J\"ulich Supercomputing Centre
  (JSC). GCS is the alliance of the three national supercomputing
  centres HLRS (Universit\"at Stuttgart), JSC (Forschungszentrum
  J\"ulich), and LRZ (Bayerische Akademie der Wissenschaften), funded
  by the German Federal Ministry of Education and Research (BMBF) and
  the German State Ministries for Research of Baden-W\"urttemberg
  (MWK), Bayern (StMWFK) and Nordrhein-Westfalen (MIWF). This work was
  supported by the Klaus Tschira foundation and the DAAD/Go8
  German-Australian exchange programme for travel support.
\end{acknowledgements}

\bibliographystyle{aa}

\begin{thebibliography}{59}
\expandafter\ifx\csname natexlab\endcsname\relax\def\natexlab#1{#1}\fi

\bibitem[{{Arnett}(1969)}]{arnett1969a}
{Arnett}, W.~D. 1969, \apss, 5, 180

\bibitem[{{Blondin} {et~al.}(2013){Blondin}, {Dessart}, {Hillier}, \&
  {Khokhlov}}]{blondin2013a}
{Blondin}, S., {Dessart}, L., {Hillier}, D.~J., \& {Khokhlov}, A.~M. 2013,
  \mnras, 429, 2127

\bibitem[{{Bulla} {et~al.}(2016){Bulla}, {Sim}, {Pakmor}, {Kromer},
  {Taubenberger}, {R{\"o}pke}, {Hillebrandt}, \& {Seitenzahl}}]{bulla2016a}
{Bulla}, M., {Sim}, S.~A., {Pakmor}, R., {et~al.} 2016, \mnras, 455, 1060

\bibitem[{{Colgate} \& {McKee}(1969)}]{colgate1969a}
{Colgate}, S.~A. \& {McKee}, C. 1969, \apj, 157, 623

\bibitem[{{Filippenko}(1997)}]{filippenko1997a}
{Filippenko}, A.~V. 1997, \araa, 35, 309

\bibitem[{Fink(2010)}]{fink2010b}
Fink, M. 2010, Dissertation, Technische Universit\"at M\"unchen

\bibitem[{{Fink} {et~al.}(2007){Fink}, {Hillebrandt}, \&
  {R{\"o}pke}}]{fink2007a}
{Fink}, M., {Hillebrandt}, W., \& {R{\"o}pke}, F.~K. 2007, \aap, 476, 1133

\bibitem[{{Fink} {et~al.}(2014){Fink}, {Kromer}, {Seitenzahl},
  {Ciaraldi-Schoolmann}, {R{\"o}pke}, {Sim}, {Pakmor}, {Ruiter}, \&
  {Hillebrandt}}]{fink2014a}
{Fink}, M., {Kromer}, M., {Seitenzahl}, I.~R., {et~al.} 2014, \mnras, 438, 1762

\bibitem[{{Fink} {et~al.}(2010){Fink}, {R{\"o}pke}, {Hillebrandt},
  {Seitenzahl}, {Sim}, \& {Kromer}}]{fink2010a}
{Fink}, M., {R{\"o}pke}, F.~K., {Hillebrandt}, W., {et~al.} 2010, \aap, 514,
  A53

\bibitem[{{Hachinger} {et~al.}(2009){Hachinger}, {Mazzali}, {Taubenberger},
  {Pakmor}, \& {Hillebrandt}}]{hachinger2009a}
{Hachinger}, S., {Mazzali}, P.~A., {Taubenberger}, S., {Pakmor}, R., \&
  {Hillebrandt}, W. 2009, \mnras, 399, 1238

\bibitem[{{Hillebrandt} {et~al.}(2013){Hillebrandt}, {Kromer}, {R{\"o}pke}, \&
  {Ruiter}}]{hillebrandt2013a}
{Hillebrandt}, W., {Kromer}, M., {R{\"o}pke}, F.~K., \& {Ruiter}, A.~J. 2013,
  Frontiers of Physics, 8, 116

\bibitem[{{Hillebrandt} \& {Niemeyer}(2000)}]{hillebrandt2000a}
{Hillebrandt}, W. \& {Niemeyer}, J.~C. 2000, \araa, 38, 191

\bibitem[{{H{\"o}flich} \& {Khokhlov}(1996)}]{hoeflich1996a}
{H{\"o}flich}, P. \& {Khokhlov}, A. 1996, \apj, 457, 500

\bibitem[{{Hoyle} \& {Fowler}(1960)}]{hoyle1960a}
{Hoyle}, F. \& {Fowler}, W.~A. 1960, \apj, 132, 565

\bibitem[{{Iben} {et~al.}(1987){Iben}, {Nomoto}, {Tornambe}, \&
  {Tutukov}}]{iben1987a}
{Iben}, Jr., I., {Nomoto}, K., {Tornambe}, A., \& {Tutukov}, A.~V. 1987, \apj,
  317, 717

\bibitem[{{Jordan} {et~al.}(2012){Jordan}, {Perets}, {Fisher}, \& {van
  Rossum}}]{jordan2012b}
{Jordan}, IV, G.~C., {Perets}, H.~B., {Fisher}, R.~T., \& {van Rossum}, D.~R.
  2012, \apjl, 761, L23

\bibitem[{{Khokhlov}(1991)}]{khokhlov1991a}
{Khokhlov}, A.~M. 1991, \aap, 245, 114

\bibitem[{{Kromer} {et~al.}(2013{\natexlab{a}}){Kromer}, {Fink}, {Stanishev},
  {Taubenberger}, {Ciaraldi-Schoolman}, {Pakmor}, {R{\"o}pke}, {Ruiter},
  {Seitenzahl}, {Sim}, {Blanc}, {Elias-Rosa}, \& {Hillebrandt}}]{kromer2013a}
{Kromer}, M., {Fink}, M., {Stanishev}, V., {et~al.} 2013{\natexlab{a}}, \mnras,
  429, 2287

\bibitem[{{Kromer} {et~al.}(2016){Kromer}, {Fremling}, {Pakmor},
  {Taubenberger}, {Amanullah}, {Cenko}, {Fransson}, {Goobar}, {Leloudas},
  {Taddia}, {R{\"o}pke}, {Seitenzahl}, {Sim}, \& {Sollerman}}]{kromer2016a}
{Kromer}, M., {Fremling}, C., {Pakmor}, R., {et~al.} 2016, \mnras, 459, 4428

\bibitem[{{Kromer} {et~al.}(2015){Kromer}, {Ohlmann}, {Pakmor}, {Ruiter},
  {Hillebrandt}, {Marquardt}, {R{\"o}pke}, {Seitenzahl}, {Sim}, \&
  {Taubenberger}}]{kromer2015a}
{Kromer}, M., {Ohlmann}, S.~T., {Pakmor}, R., {et~al.} 2015, \mnras, 450, 3045

\bibitem[{{Kromer} {et~al.}(2013{\natexlab{b}}){Kromer}, {Pakmor},
  {Taubenberger}, {Pignata}, {Fink}, {R{\"o}pke}, {Seitenzahl}, {Sim}, \&
  {Hillebrandt}}]{kromer2013b}
{Kromer}, M., {Pakmor}, R., {Taubenberger}, S., {et~al.} 2013{\natexlab{b}},
  \apjl, 778, L18

\bibitem[{{Kromer} \& {Sim}(2009)}]{kromer2009a}
{Kromer}, M. \& {Sim}, S.~A. 2009, \mnras, 398, 1809

\bibitem[{{Kromer} {et~al.}(2010){Kromer}, {Sim}, {Fink}, {R{\"o}pke},
  {Seitenzahl}, \& {Hillebrandt}}]{kromer2010a}
{Kromer}, M., {Sim}, S.~A., {Fink}, M., {et~al.} 2010, \apj, 719, 1067

\bibitem[{{Magee} {et~al.}(2016){Magee}, {Kotak}, {Sim}, {Kromer},
  {Rabinowitz}, {Smartt}, {Baltay}, {Campbell}, {Chen}, {Fink}, {Gal-Yam},
  {Galbany}, {Hillebrandt}, {Inserra}, {Kankare}, {Le Guillou}, {Lyman},
  {Maguire}, {Pakmor}, {R{\"o}pke}, {Ruiter}, {Seitenzahl}, {Sullivan},
  {Valenti}, \& {Young}}]{magee2016a}
{Magee}, M.~R., {Kotak}, R., {Sim}, S.~A., {et~al.} 2016, \aap, 589, A89

\bibitem[{{Maoz} {et~al.}(2014){Maoz}, {Mannucci}, \& {Nelemans}}]{maoz2014a}
{Maoz}, D., {Mannucci}, F., \& {Nelemans}, G. 2014, \araa, 52, 107

\bibitem[{{Marquardt} {et~al.}(2015){Marquardt}, {Sim}, {Ruiter}, {Seitenzahl},
  {Ohlmann}, {Kromer}, {Pakmor}, \& {R{\"o}pke}}]{marquardt2015a}
{Marquardt}, K.~S., {Sim}, S.~A., {Ruiter}, A.~J., {et~al.} 2015, \aap, 580,
  A118

\bibitem[{{Moll} {et~al.}(2014){Moll}, {Raskin}, {Kasen}, \&
  {Woosley}}]{moll2014a}
{Moll}, R., {Raskin}, C., {Kasen}, D., \& {Woosley}, S.~E. 2014, \apj, 785, 105

\bibitem[{{Moll} \& {Woosley}(2013)}]{moll2013a}
{Moll}, R. \& {Woosley}, S.~E. 2013, \apj, 774, 137

\bibitem[{{Nomoto} \& {Iben}(1985)}]{nomoto1985a}
{Nomoto}, K. \& {Iben}, Jr., I. 1985, \apj, 297, 531

\bibitem[{{Ohlmann} {et~al.}(2014){Ohlmann}, {Kromer}, {Fink}, {Pakmor},
  {Seitenzahl}, {Sim}, \& {R{\"o}pke}}]{ohlmann2014a}
{Ohlmann}, S.~T., {Kromer}, M., {Fink}, M., {et~al.} 2014, \aap, 572, A57

\bibitem[{{Pakmor} {et~al.}(2012{\natexlab{a}}){Pakmor}, {Edelmann},
  {R{\"o}pke}, \& {Hillebrandt}}]{pakmor2012b}
{Pakmor}, R., {Edelmann}, P., {R{\"o}pke}, F.~K., \& {Hillebrandt}, W.
  2012{\natexlab{a}}, \mnras, 424, 2222

\bibitem[{{Pakmor} {et~al.}(2011){Pakmor}, {Hachinger}, {R{\"o}pke}, \&
  {Hillebrandt}}]{pakmor2011b}
{Pakmor}, R., {Hachinger}, S., {R{\"o}pke}, F.~K., \& {Hillebrandt}, W. 2011,
  \aap, 528, A117+

\bibitem[{{Pakmor} {et~al.}(2010){Pakmor}, {Kromer}, {R{\"o}pke}, {Sim},
  {Ruiter}, \& {Hillebrandt}}]{pakmor2010a}
{Pakmor}, R., {Kromer}, M., {R{\"o}pke}, F.~K., {et~al.} 2010, \nat, 463, 61

\bibitem[{{Pakmor} {et~al.}(2012{\natexlab{b}}){Pakmor}, {Kromer},
  {Taubenberger}, {Sim}, {R{\"o}pke}, \& {Hillebrandt}}]{pakmor2012a}
{Pakmor}, R., {Kromer}, M., {Taubenberger}, S., {et~al.} 2012{\natexlab{b}},
  \apjl, 747, L10

\bibitem[{{Pakmor} {et~al.}(2013){Pakmor}, {Kromer}, {Taubenberger}, \&
  {Springel}}]{pakmor2013a}
{Pakmor}, R., {Kromer}, M., {Taubenberger}, S., \& {Springel}, V. 2013, \apjl,
  770, L8

\bibitem[{{Perlmutter} {et~al.}(1999){Perlmutter}, {Aldering}, {Goldhaber},
  {Knop}, {Nugent}, {Castro}, {Deustua}, {Fabbro}, {Goobar}, {Groom}, {Hook},
  {Kim}, {Kim}, {Lee}, {Nunes}, {Pain}, {Pennypacker}, {Quimby}, {Lidman},
  {Ellis}, {Irwin}, {McMahon}, {Ruiz-Lapuente}, {Walton}, {Schaefer}, {Boyle},
  {Filippenko}, {Matheson}, {Fruchter}, {Panagia}, {Newberg}, {Couch}, \& {The
  Supernova Cosmology Project}}]{perlmutter1999a}
{Perlmutter}, S., {Aldering}, G., {Goldhaber}, G., {et~al.} 1999, \apj, 517,
  565

\bibitem[{{Phillips}(1993)}]{phillips1993a}
{Phillips}, M.~M. 1993, \apjl, 413, L105

\bibitem[{{Plewa} {et~al.}(2004){Plewa}, {Calder}, \& {Lamb}}]{plewa2004a}
{Plewa}, T., {Calder}, A.~C., \& {Lamb}, D.~Q. 2004, \apjl, 612, L37

\bibitem[{{Reinecke} {et~al.}(2002){Reinecke}, {Hillebrandt}, \&
  {Niemeyer}}]{reinecke2002b}
{Reinecke}, M., {Hillebrandt}, W., \& {Niemeyer}, J.~C. 2002, \aap, 386, 936

\bibitem[{{Riess} {et~al.}(1998){Riess}, {Filippenko}, {Challis},
  {Clocchiatti}, {Diercks}, {Garnavich}, {Gilliland}, {Hogan}, {Jha},
  {Kirshner}, {Leibundgut}, {Phillips}, {Reiss}, {Schmidt}, {Schommer},
  {Smith}, {Spyromilio}, {Stubbs}, {Suntzeff}, \& {Tonry}}]{riess1998a}
{Riess}, A.~G., {Filippenko}, A.~V., {Challis}, P., {et~al.} 1998, \aj, 116,
  1009

\bibitem[{{R{\"o}pke} {et~al.}(2012){R{\"o}pke}, {Kromer}, {Seitenzahl},
  {Pakmor}, {Sim}, {Taubenberger}, {Ciaraldi-Schoolmann}, {Hillebrandt},
  {Aldering}, {Antilogus}, {Baltay}, {Benitez-Herrera}, {Bongard}, {Buton},
  {Canto}, {Cellier-Holzem}, {Childress}, {Chotard}, {Copin}, {Fakhouri},
  {Fink}, {Fouchez}, {Gangler}, {Guy}, {Hachinger}, {Hsiao}, {Chen},
  {Kerschhaggl}, {Kowalski}, {Nugent}, {Paech}, {Pain}, {Pecontal}, {Pereira},
  {Perlmutter}, {Rabinowitz}, {Rigault}, {Runge}, {Saunders}, {Smadja},
  {Suzuki}, {Tao}, {Thomas}, {Tilquin}, \& {Wu}}]{roepke2012a}
{R{\"o}pke}, F.~K., {Kromer}, M., {Seitenzahl}, I.~R., {et~al.} 2012, \apjl,
  750, L19

\bibitem[{{Saio} \& {Nomoto}(2004)}]{saio2004a}
{Saio}, H. \& {Nomoto}, K. 2004, \apj, 615, 444

\bibitem[{{Scalzo} {et~al.}(2014){Scalzo}, {Aldering}, {Antilogus}, {Aragon},
  {Bailey}, {Baltay}, {Bongard}, {Buton}, {Cellier-Holzem}, {Childress},
  {Chotard}, {Copin}, {Fakhouri}, {Gangler}, {Guy}, {Kim}, {Kowalski},
  {Kromer}, {Nordin}, {Nugent}, {Paech}, {Pain}, {Pecontal}, {Pereira},
  {Perlmutter}, {Rabinowitz}, {Rigault}, {Runge}, {Saunders}, {Sim}, {Smadja},
  {Tao}, {Taubenberger}, {Thomas}, {Weaver}, \& {Nearby Supernova
  Factory}}]{scalzo2014a}
{Scalzo}, R., {Aldering}, G., {Antilogus}, P., {et~al.} 2014, \mnras, 440, 1498

\bibitem[{{Seitenzahl} {et~al.}(2013){Seitenzahl}, {Ciaraldi-Schoolmann},
  {R{\"o}pke}, {Fink}, {Hillebrandt}, {Kromer}, {Pakmor}, {Ruiter}, {Sim}, \&
  {Taubenberger}}]{seitenzahl2013a}
{Seitenzahl}, I.~R., {Ciaraldi-Schoolmann}, F., {R{\"o}pke}, F.~K., {et~al.}
  2013, \mnras, 429, 1156

\bibitem[{{Seitenzahl} {et~al.}(2016){Seitenzahl}, {Kromer}, {Ohlmann},
  {Ciaraldi-Schoolmann}, {Marquardt}, {Fink}, {Hillebrandt}, {Pakmor},
  {R{\"o}pke}, {Ruiter}, {Sim}, \& {Taubenberger}}]{seitenzahl2016a}
{Seitenzahl}, I.~R., {Kromer}, M., {Ohlmann}, S.~T., {et~al.} 2016, \aap, 592,
  A57

\bibitem[{{Shen} {et~al.}(2012){Shen}, {Bildsten}, {Kasen}, \&
  {Quataert}}]{shen2012a}
{Shen}, K.~J., {Bildsten}, L., {Kasen}, D., \& {Quataert}, E. 2012, \apj, 748,
  35

\bibitem[{{Shen} {et~al.}(2010){Shen}, {Kasen}, {Weinberg}, {Bildsten}, \&
  {Scannapieco}}]{shen2010a}
{Shen}, K.~J., {Kasen}, D., {Weinberg}, N.~N., {Bildsten}, L., \&
  {Scannapieco}, E. 2010, \apj, 715, 767

\bibitem[{{Shigeyama} {et~al.}(1992){Shigeyama}, {Nomoto}, {Yamaoka}, \&
  {Thielemann}}]{shigeyama1992a}
{Shigeyama}, T., {Nomoto}, K., {Yamaoka}, H., \& {Thielemann}, F. 1992, \apjl,
  386, L13

\bibitem[{{Sim} {et~al.}(2012){Sim}, {Fink}, {Kromer}, {R{\"o}pke}, {Ruiter},
  \& {Hillebrandt}}]{sim2012a}
{Sim}, S.~A., {Fink}, M., {Kromer}, M., {et~al.} 2012, \mnras, 420, 3003

\bibitem[{{Sim} {et~al.}(2010){Sim}, {R{\"o}pke}, {Hillebrandt}, {Kromer},
  {Pakmor}, {Fink}, {Ruiter}, \& {Seitenzahl}}]{sim2010a}
{Sim}, S.~A., {R{\"o}pke}, F.~K., {Hillebrandt}, W., {et~al.} 2010, \apjl, 714,
  L52

\bibitem[{{Sim} {et~al.}(2013){Sim}, {Seitenzahl}, {Kromer},
  {Ciaraldi-Schoolmann}, {R{\"o}pke}, {Fink}, {Hillebrandt}, {Pakmor},
  {Ruiter}, \& {Taubenberger}}]{sim2013a}
{Sim}, S.~A., {Seitenzahl}, I.~R., {Kromer}, M., {et~al.} 2013, \mnras, 436,
  333

\bibitem[{Stephan \& Docter(2015)}]{stephan2015a}
Stephan, M. \& Docter, J. 2015, Journal of large-scale research facilities
  {JLSRF}, 1, A1

\bibitem[{{Taam}(1980)}]{taam1980a}
{Taam}, R.~E. 1980, \apj, 242, 749

\bibitem[{{Tanikawa} {et~al.}(2015){Tanikawa}, {Nakasato}, {Sato}, {Nomoto},
  {Maeda}, \& {Hachisu}}]{tanikawa2015a}
{Tanikawa}, A., {Nakasato}, N., {Sato}, Y., {et~al.} 2015, \apj, 807, 40

\bibitem[{{Travaglio} {et~al.}(2004){Travaglio}, {Hillebrandt}, {Reinecke}, \&
  {Thielemann}}]{travaglio2004a}
{Travaglio}, C., {Hillebrandt}, W., {Reinecke}, M., \& {Thielemann}, F.-K.
  2004, \aap, 425, 1029

\bibitem[{{Webbink}(1984)}]{webbink1984a}
{Webbink}, R.~F. 1984, \apj, 277, 355

\bibitem[{{Whelan} \& {Iben}(1973)}]{whelan1973a}
{Whelan}, J. \& {Iben}, I.~J. 1973, \apj, 186, 1007

\bibitem[{{Woosley} \& {Kasen}(2011)}]{woosley2011b}
{Woosley}, S.~E. \& {Kasen}, D. 2011, \apj, 734, 38

\bibitem[{{Yaron} \& {Gal-Yam}(2012)}]{yaron2012a}
{Yaron}, O. \& {Gal-Yam}, A. 2012, \pasp, 124, 668

\end{thebibliography}

\end{document}